\documentstyle[aaspp4]{article}

\def\mincir{\ \raise -2.truept\hbox{\rlap{\hbox{$\sim$}}\raise5.truept  
\hbox{$<$}\ }}
\def\magcir{\ \raise -2.truept\hbox{\rlap{\hbox{$\sim$}}\raise5.truept  
\hbox{$>$}\ }}

\def\ie{i.e.\ }
\def\eg{e.g.\ }
\def\cf{{\it cf.\/} }
\def\be{\begin{equation}}
\def\ee{\end{equation}}
\def\bdm{\begin{displaymath}}
\def\edm{\end{displaymath}}
\def\ea{{et al.\/} }

\def\kmpersec{km s$^{-1}$}
\def\lta{\mathrel{\rlap{\lower 3pt\hbox{$\mathchar"218$}}
    \raise 2.0pt\hbox{$\mathchar"13C$}}}

\newbox\grsign \setbox\grsign=\hbox{$>$} \newdimen\grdimen \grdimen=\ht\grsign
\newbox\simlessbox \newbox\simgreatbox
\setbox\simgreatbox=\hbox{\raise.5ex\hbox{$>$}\llap
     {\lower.5ex\hbox{$\sim$}}}\ht1=\grdimen\dp1=0pt
\setbox\simlessbox=\hbox{\raise.5ex\hbox{$<$}\llap
     {\lower.5ex\hbox{$\sim$}}}\ht2=\grdimen\dp2=0pt
\newcommand{\simgt}{\mathrel{\copy\simgreatbox}}
\newcommand{\simlt}{\mathrel{\copy\simlessbox}}

\tighten

\begin{document}

\title{Dependence of Halo Properties on Interaction
History, Environment and Cosmology}

\author{Jeffrey P. Gardner$^{1,2}$}
\affil{University of Pittsburgh, Department of Physics and Astronomy,
Pittsburgh, PA, 15260}
\footnotetext[1]
{Institute of Astronomy, Madingley Road, Cambridge, CB3 0HA, UK}
\footnotetext[2]
{NSF-NATO Postdoctoral Fellow}

\begin{abstract}

I present results from numerical N-body simulations regarding the
effect of merging events on the angular momentum distribution of
galactic halos as well as a comparison of halo growth in Press-Schechter
vs.\ N-body methods.  A total of six simulations are used spanning 3
cosmologies: a standard flat $\Omega_0=1$ model, an open
$\Omega_0=0.3$ model and a ``tilted'' flat $\Omega_0=1$ model with
spectral index $n=0.8$.  In each model, one run was conducted using a
spatially uniform grid of particles and one using a refined grid in a large
void.  In all three models and all environments tested, the mean
angular momentum of merger remnants (halo interaction products with
mass ratios 3:1 or less) is greater than non-merger remnants.
Furthermore, the dispersion in the merger-remnant angular momentum
distribution is smaller than the dispersion of the non-merger
distribution.  The interpretation most consistent with the data is
that the orbital angular momentum of the interactors is important in
establishing the final angular momentum of the merger product.  I give the
angular momentum distribution which describes the merger remnant population.

I trace the most massive progenitor of $L_*$ galactic-mass halos
(uniform grid) and $10^{11} M_\odot$ halos (refined void) from
redshift $z=0$ back to $z=5$.  Monte-Carlo mass histories
match simulations reasonably well for the latter sample.  I find that
for halos of mass $10^{12} \simlt M \simlt 10^{14} M_\odot$, this
method can underestimate the mass of progenitors by 20\%, hence
yielding improper formation redshifts of halos.  With this caveat,
however, the general shapes of halo mass histories and formation-time
distributions are preserved.

\end{abstract}

\section{Introduction}

The majority of spiral galaxies do not exist in isolation, but rather
in an environment which sustains repeated close encounters and merging
processes with both small and large companions (\cite{SD00}).
Increasing evidence links disturbed or lopsided features of these
galaxies to recent dynamical interactions.  HST observations at high
redshift have shown a preponderance of ``train wreck'' morphologies
arising from major interactions.  More recently, minor mergers have
begun to be linked to $m=1$ mode disturbances as observed in the
Fourier azimuthal decomposition of the galaxy's surface brightness,
generally denoted as ``lopsidedness'' (\cite{RRK00}; \cite{RR98}).
Specifically, studies by Richter \& Sancisi (1994) have estimated that about
half of late-type spirals exhibit
significantly lopsided HI profiles, and half of these are field
galaxies removed from cluster environments.  Furthermore,
Zaritsky \& Rix find that 20\% of all late-type spirals are significantly
lopsided at optical and near-infrared
wavelengths, yet have no obvious interactors.  Rudnick \& Rix (1998) find that
one fifth of all disk galaxies, regardless of Hubble type, have
azimuthal asymmetries in their stellar mass distribution.  They also show that
this observed asymmetry is in fact due to the underlying mass
distribution and not dust obscuration or other non-dynamical effects.
Dynamical profiles of this nature have been shown to be consistent
with the accretion of a satellite galaxy of a larger system
(\cite{LS98}; \cite{SD00}; and references therein).
Consequently, it is probable that a substantial fraction of disk
galaxies are experiencing interaction due to minor mergers.

Field spheroidals also have a number of properties such as shells,
counterrotating inner disk, and bluer colors, which are believed to
also be associated with recent merger events (\cite{Decarvahlo92};
\cite{Rose94}; \cite{Longhetti98}; \cite{Abraham98}).  By counting
major mergers in large-scale simulations, Governato \ea (1999) show
that a substantial fraction of the field elliptical population could
be the result of major merging activity.

Hence, even in the field, observational evidence exists to support the
hierarchical clustering scenario (HCS, see Peebles 1993) wherein larger
objects are formed through the progressive assemblage of smaller mass
collapsed structures.  In short, the galaxy formation and evolution
process is ubiquitously influenced by interactions.  Therefore, to
fully understand galactic formation and evolution and their consequences
(both observational and theoretical), one must understand link between
galaxy properties and interaction history, and how both of these may
be modified by environmental effects.

The Press \& Schechter formalism (1974, PS) is an
analytical model to describe the hierarchical clustering scenario.
The PS theory allows one to derive, in a
relatively simple and straightforward way, the mass distribution of
collapsed objects and its evolution with time in different FRW
universes.

Despite its simple derivation, which relies uniquely on the linear
theory, the PS formula captures the major feature of the hierarchical
evolution as shown in the good fit of the PS mass function to N-body
simulations (\cite{frenk88}; \cite{lc94}; \cite{GKHW}; \cite{brtn99}
and references therein).  Recently it has become possible to pinpoint
quantitative differences between this picture and the more complex
N--body description.  At cluster-sized masses, it is possible to match
PS and simulated mass functions reasonably well
(\cite{governatocluster99}; \cite{brtn99}), albeit with some tuning.
However, at lower masses, evidence suggests that PS may provide a
systematic overestimate of the halo number density (\cite{bn98};
\cite{gross98}; \cite{tormen98}).

The PS formalism can be extended to give the probability that a halo
of mass $M_0$ at redshift $z_0$ was also in a halo of mass $M_1$ at
time $z_1$ (\cite{bcek91}). This Extended Press-Schechter (EPS) can be
further instrumented to construct halo formation times, mass
histories, and merger rates in Monte-Carlo techniques (\cite{lc93}).
Additional enhancements can be made which follow the detailed
evolution and interaction of halos (\cite{baugh98} and references
therein) in the form of semi-analytic merger trees (SAM) which are far
less costly than N-body simulations.  The SAM formalism has been used
to predict a wide range of galaxy properties such as colors,
morphology, merging history, and formation time (\cite{baugh96} and
references therein).  It has been used successfully to predict the
same qualitative behavior found in N-body simulations.  For instance,
both SAM studies and N-body simulations find that the typical redshift
of last major interaction of field ellipticals is lower than the
cluster elliptical population (\cite{mergerpaper}; \cite{baugh96}).
Furthermore, SAM results also generally reproduce observed behavior
such as the morphology-density relation, color-magnitude diagrams, the
Butcher-Oemler effect, and the luminosity evolution of color-selected
galaxy samples (\cite{kauffmann95}; \cite{kauffmann96}; \cite{kc98a};
\cite{kc98b})

Given the great utility of PS methods, it is crucial to
understand how the PS expression of structure formation relates to
N-body simulations.  PS predictions began to be tested against N-body
results a number of years ago (\cite{efstathiou88}; \cite{lc94}) and
generally found, within the dynamic range of the simulations, to
compare relatively well, although soon results were found that hinted
at discrepancies between PS and simulations (\cite{gb94}).  As the
range in masses accessible in the N-body method has increased, a
number of other studies have been performed (\cite{somerville00};
\cite{tormen98}; \cite{sheth99}; \cite{brtn99};
\cite{governatocluster99}; \cite{grammann99}; \cite{bn98};
\cite{gross98}).  Somerville \ea (2000) compare the extended
Press-Schechter model and find that, at high masses, discrepancies as
high as 50\% can exist between SAM and N-body predictions of
progenitor numbers.  They find discrepancies in such quantities as the
PS mass function and conditional mass function, as well as SAM
number-mass distributions.  On the other hand, they find that a number
of more relative quantities, such as the ratio of progenitor masses,
are reproduced fairly well.  Tormen (1998) also finds that EPS
underpredicts the number of high-mass progenitors of cluster members at
high redshift and concludes that caution should be exercised when
modeling clustering properties of halo progenitors.  

PS provides a representation of halos over a cosmologically
representative range of environment.  How applicable is the PS mass function in
environmental extremes?  Lemson \& Kauffmann (1999) use N-body
simulations to examine a number of halo parameters as function of
environment.  They find environmental dependence only for the mass
distribution of halos, with angular momentum, concentration parameter,
and formation time remaining unaffected.  Consequently, one would
expect PS to maintain as accurate a representation of environmental
extremes as it does more representative regimes.  In this work, I
compare PS output to N-body simulations in both a representative
volume and an extremely underdense region.

One quantity which cannot be derived from semi-analytic modeling is
the angular momentum properties of a halo.  Workers who study
structure within galactic halos may derive halo mass and interaction
histories from PS based methods, but spin parameters are typically
assigned from the prescriptions found by previous N-body work.  For
example, Dalcanton, Spergel, \& Summers (1997) use PS in combination
with the spin parameter distribution of Warren \ea (1992) in order to
study the distributions of surface brightness and scale length in disk
galaxies.  Van den Bosch (1998) uses this same distribution to examine
the formation of disks and bulges within dark matter halos.  Mo, Mao
\& White (1998) examine the formation of disk galaxies by
incorporating an {\it a priori} distribution of dark matter halo
angular momentum.  Any SAM investigation of galaxy evolution and
interaction (e.g. \cite{sp99}) must also assign spin parameters in
this manner.  When a halo experiences a major merging event, a new
spin parameter can be assigned.  However, should it be taken from the
same general distribution found in N-body simulations or are the
angular momenta of merger remnants systematically different from the
quiescent population?

This paper is divided into two major thrusts.  In the first, I examine
the distribution of halo spin parameters in the overall halo
population as well as merger remnants.  I do this for several
cosmologies, and for a large cosmologically representative volume
(``uniform volume'') as well as a large, underdense void.
In the second thrust, I compare semi-analytic halo progenitor masses
and halo formation times to N-body simulations.  I examine this
relation in 3 uniform volumes and two voids, which collectively sample
3 different cosmologies.

\section{Methods}

\subsection{The PS formalism}

The standard PS formula predicts the number density $N(M,z)\, dM$ of
high contrast structure of a given mass $M$ at a given epoch.  The
counting of such structures is based on the hypothesis that the
overdense regions in the initial density field filtered on the scale
of interest, will eventually collapse to form high contrast structure.
Then to obtain the fraction of the mass collapsed on a certain scale,
it needs to single out volumes of the filtered density field, where
the (linearly) evolved density contrast $\delta $ exceeds a nominal
threshold $\delta_c$ depending on the epoch, which guarantees that
the associated volume is collapsed into a virialized object.  The PS
formula states:
\begin{equation}
\label{eq:PS}
N(M,z)~ dM=\sqrt{2\over \pi}~ {\rho_0\over M}~ {\delta_c \over \sigma^2(M) }
e^{\delta_c^2/2\sigma^2(M)}~ {d\sigma(M)\over dM}~ dM ,
\end{equation}
where $\sigma^2(M)$ is the mass variance of the perturbation field on
the mass scale $M$:
\begin{equation}
\sigma^2(M) = {1\over 2 \pi^2} \int P(k) W^2(kR) k^2 dk,
\end{equation}
where $P(k)$ is the power spectrum and $W(kR)$ the Fourier transform
if the real-space top-hat function.  The conventional way to normalize
the power spectrum is $\sigma_8$, the rms value of the linear
fluctuation in the mass distribution on scales of $R=8\, h^{-1}$ Mpc.
The field is taken to be Gaussian, and then the variance
$S=\sigma^2(M)$ completely describes the density perturbation field.
The threshold $\delta_c$ refers to the collapse of homogeneous
spherical volumes and is found to be $\delta_c(z=0)=1.68$ for
$\Omega_0=1$ models.  The time variation of $\delta_c(z)$ is
proportional to the linear growth factor $D(t)$ (see e.g. Peebles
1993) and contains the dependence from the specific FRW model.  The
overall normalization is such that at any $z$ all the available mass
density $\rho$ resides in collapsed objects.  For a detailed
derivation and the analysis of the underlying statistics, see Bond et
al (1991).

Paulo Tozzi graciously furnished the author with a copy of his code,
an implementation of the Monte-Carlo method given in Lacey \& Cole
(1993) which is in turn based on the EPS ``excursion set'' formalism
developed by Bond \ea (1991).  In this method, the linear density
field is smoothed at progressively larger mass scales and the mass of
the halo containing a given particle at time $t_1$ is equal to the
mass $M_1$ of the largest sphere which in which the average
overdensity $\delta_c(t_1)$ exceeds the collapse threshold.  This
ensures that a halos of mass $M$ has not yet been subsumed by a larger
mass halo.  $\delta_c(t)$, of course, changes with time and thus at
later times, our halo of mass $M_1$ and time $t_1$ may ultimately
become part of a larger collapsed structure $M_2$ which does not reach
critical density until time $t_2$.  By examining the growth of linear
modes in the same region, one can statistically construct halo growth
and interaction histories.  Tozzi's method differs from that of Lacey
\& Cole (1993) in the specifics of how it calculates the relationship
between mass ($M_1, M_2$, \ldots) and collapse threshold
($\delta_c(t_1), \delta_c(t_2)$, \ldots) but yields essentially the
same results (Tozzi, private communication).  The method is described
in detail by Governato \& Tozzi (2001).

\subsection{Simulations}

I present 6 simulations which are evolved from 3 cosmological models:
a critical universe (SCDM) ($\Omega_0=1$, $h\equiv H_0/100$ \kmpersec\
Mpc$^{-1}=0.5$, $\sigma_8 =0.7$), an open (OCDM) universe
($\Omega_0=0.3$, $h=0.75$, $\sigma_8 =1$) and a tilted critical model
(TCDM) with h$=0.5$, $\sigma_8 =0.6$, a primordial index n$=0.8$ and a
gamma factor $\Gamma =$ 0.37.  The transfer function of Bardeen \ea
(1986) was used in all cases.  The simulated volume was 100 Mpc on a
side (h already included) in all six runs.  The parameters of the TCDM
model have been chosen to satisfy both the cluster abundance and the
COBE normalization at very large scales (\cite{cmt98}).  This model
differs from the SCDM in having less power and a steeper power
spectrum at scales under 8h$^{-1}$ Mpc in comparison with the other
two models.  Each simulation was performed using PKDGRAV (Stadel \&
Quinn, in preparation) a parallel N--body treecode supporting periodic
boundary conditions.  All runs were started at redshifts sufficiently
high to ensure that the absolute maximum density contrast $|\delta|
\simlt 1$.  All simulations were evolved using between 500 and 1000
major time steps in which the forces on all particles were calculated.
Each major time step is divided into a number of smaller substeps, the
exact number of which is adaptively chosen by PKDGRAV such that all
particles are moved on steps consonant with their dynamical times.

One simulation in each cosmology (the ``uniform volume'') was run
beginning from a spatially uniform grid of $144^3$ ($\sim$ 3 million)
equal--mass particles with spline softening set to 60 kpc, allowing us
to resolve individual halos with present-day circular velocities $V_c$
as low as 130 \kmpersec\ with 100 particles in OCDM and 170 \kmpersec\
with 100 particles in SCDM and TCDM.  The particle masses were $1.57
\times 10^{10}$ M$_\odot$ in OCDM and $2.32 \times 10^{10}$ M$_\odot$
in SCDM and TCDM, while the opening angles of the SCDM and OCDM
simulations were $\theta=0.5$ for redshifts $z>2$ and $\theta=0.7$
thereafter.  TCDM was run with $\theta=0.5$ at all redshifts.  

The remaining three simulations, one in each cosmology, were of a
large void approximately 40 Mpc in diameter, where I defined the void
region as the largest sphere that could be drawn in the simulation
volume with mean overdensity $\bar\delta = -0.9$.  The radii of the
resultant spheres were 18.5 Mpc, 17 Mpc, and 13.5 Mpc for SCDM, OCDM
and TCDM cosmologies respectively.  The void runs were performed using
the hierarchical grids technique in which the void region is
represented with large numbers of low-mass particles, while the
surrounding environment is composed of small numbers of high-mass
particles.  This allows much greater mass resolution to be achieved
while ensuring that cosmological context is maintained.  I used
approximately 5 million particles in each void simulation, with an
effective resolution of $432^3$ across the 100 Mpc box.  $\theta$ was
set to 0.6 for $z>2$ and 0.8 thereafter.  Particles were
spline-softened to 6.67 kpc, being $8.6\times 10^8$ M$_\odot$ in mass
in critical models and $5.8\times 10^8$ in OCDM.  Thus, structure in
the voids was resolved in mass at a factor of 27 superior to the
uniform volume runs.  For convenience, I shall adopt the notation of
SCDMu, OCDMu and TCDMu for the uniform volumes and SCDMv, OCDMv and
TCDMv for the voids.

Throughout the paper, I identify halos using two primary methods.  The
first is the classic friends-of-friends (FOF) method (\cite{fofpaper})
using a linking length that corresponds to the mean interparticle
separation at the density contour defining the virial radius of an
isothermal sphere: $b=(n \rho_{vir}(z) / \bar\rho(z) / 3.0)^{-1/3}$
where $n$ is the particle number density and $\rho_{vir}(z)$ is the
virial density which depends on cosmology and is given in Kitayama \&
Suto (1996).  The second is a variant of the spherical overdensity
(SO) technique (\cite{lc94}) where I determine the most bound particle
in a FOF group and find the radius from that particle at which the enclosed
density crosses the virial threshold $\rho_{vir}(z)$.

\subsection{Halo Mass Histories}
\label{ssec:masshistmethod}

In the simplest test between PS and N--body simulations, one assumes as input
the spectral parameter $P(k)$ and $\sigma_8$, and compares the mass
distribution at a given epoch.  This essentially decides whether the
PS formula correctly states the shapes and the normalization of the
mass distribution of collapsed objects.  Work has been done comparing
the PS mass function to N-body simulations (\cite{lc94},
\cite{brtn99}, \cite{governatocluster99}, \cite{bn98}, \cite{gross98},
\cite{tormen98}), and I do not use this method in my analysis.

I focus on another test which examines the evolution of the linear
density field and compares it to the hierarchical clustering paradigm
(\cf \cite{kaiser86}) as realized in the N-body method.  The parameter
$\delta_c$ (\cf equation (\ref{eq:PS})) is the value of linear
overdensity above which a perturbation is considered to be collapsed
(evolved to the present day).  Thus, $\delta_c$ determines the mapping
between the linear perturbation field and the mass of a halo.  By
demanding that the $z=0$ PS mass function matches the z=0 mass
function of the simulation, one is effectively anchoring $\delta_c$.
This corresponds to the method used by Tormen (1998) in examining the
conditional mass function of the calibrated EPS method.  Thus there are
no remaining free parameters to the evolution of the density
perturbation field unless one invokes a $\delta_c(z)$ which varies
non-linearly with redshift such as the one examined by Governato et al
(1999).

In the uniform volume simulations, I identified all halos at z=0 of at
least mass $M_{res}=2.318\times 10^{12}$ M$_\odot$, which corresponds
to 100 particles in the flat models and roughly 150 in OCDM.  With a
mass-based cutoff, it is more straightforward to compare time
evolution of the different cosmologies.  Having at least 100 particles
also insures that the halos are substantially larger than their
softening lengths, guarding against the influence of numerical effects
on global halo properties.  I then ``traced'' each identified halo
backwards in time, following the mass of the most massive progenitor
as a function of redshift.  In the case of the void runs, I identified
all groups with at least 50 particles and traced them in a manner
identical to the uniform volume counterparts.  Unfortunately, the TCDM
cosmology forms structure extremely late, and the void had only 12
halos in it greater than this mass cutoff.  I therefore drop TCDMv
from the following analysis.

I now wish to construct a Monte-Carlo realization which most
accurately compares against the simulated results.  Comparing
individual halos is fruitless, given the extreme variation of mass
histories even between halos of the same mass.  Instead I wish to
characterize the mean mass history of the sample, \ie the average mass
of a halo in the sample as a function of redshift.  This was easily
done by drawing the mass distribution of the halos to be modeled
from the simulations themselves.  Specifically, for
each $M>M_{res}$ halo in a simulation, 100 Monte-Carlo
realizations of the mass history of that halo were constructed.  Thus,
I can directly compare the aggregate mass history of the sample of
simulated halos to the collective mass history of the Monte-Carlo
realizations.  

Errors in both simulated and Monte-Carlo samples were estimated
using the bootstrap method.  For each sample, 1000 identically-sized
subsamples of halos were drawn from the z=0 distribution.  Then, for
each subsample, the mean mass history was calculated and used in the
bootstrap estimate.

\subsection{Halo Angular Momentum}

For SCDMu and OCDMu uniform volumes and for the OCDMv void, I
determine the distribution of halo spin parameters $\lambda =
LE^{1/2}/GM^{5/2}$ for recent merger remnants versus non-remnants,
where $E$ is the total energy of the halo, $L$ the angular momentum,
and $M$ the mass.  I select halos for this analysis based on particle
number in the $z=0$ FOF group: 100 particles for SCDMu and OCDMu and
50 particles for OCDMv.  The OCDMu particle cutoff is lower than in
the section \ref{ssec:masshistmethod} for more favorable number
statistics.  To identify mergers, I use the groups determined by SKID,
a halo-finding algorithm based on local density maxima.  The SKID
algorithm is similar to the DENMAX scheme (Gelb \& Bertschinger, 1994)
as it groups particles by moving them along the density gradient to
the local density maximum.  The density field and density gradient is
defined everywhere by smoothing each particle with a cubic spline
weighting function of size determined by the distance encompassing the
nearest 32 neighbors.  At a given redshift, only particles with local
densities greater than one-third of 177 times the critical density are
``skidded'' to the local density maximum.  This threshold corresponds
roughly to the local density at the virial radius. The final step of
the process is to remove all particles that are not gravitationally
bound to their parent halo.  SKID was originally designed to find high
contrast density structures within larger halos (see \cite{ghigna98}
for a more complete description).  For halos with no resolved internal
substructure (i.e. no halos within halos) it gives results very
similar to FOF. However, SKID does not suffer of the well known
pathology of FOF of linking together close binary systems
(\cite{governato97}).  Instead, it only combines two ``clumps'' of
substructure when they are mutually gravitationally bound and thus is
well suited for the task of identifying binary systems in the process
of merging.

I denote a halo as a merger remnant if, at some time during $0<z<z_m$
it was classified as a single SKID group in one output, but two
separate groups with a mass ratio $\leq 3:1$ in the preceding output.
This method is essentially independent of output spacing except in the
extreme case where the halo mass does grows quiescently by order its
own mass between outputs.  $z_m$ was chosen to be as close to $z=0.5$
as the outputs would allow.  In SCDMu this is $z_m=0.4$, and in both
OCDM runs $z_m=0.5$.

I use SKID to identify halos before the merging event and spherical
overdensity to identify the descendant halo at the present time.  The
SO method has the advantage of culling out pathologically shaped
groups which depart substantially from the isothermal sphere model.
The most common example is a ``dumbbell'' halo, which consists of two
spherical halos that FOF links together by a thin bridge of particles
spaced close to $\rho_{vir}(z)/3$.  The total spin parameter of such a
configuration is irrelevant to this study which is concerned only with
cases where the angular momentum of a galaxy is linked to its parent
halo.  The spin parameter $\lambda$ is calculated by considering the
contributions of all particles within the SO virial radius.  If I use
the FOF group membership rather than the SO, the angular momenta for
non-pathological cases (about the FOF center of mass) is generally the
same.  Strangely-shaped groups are in the minority and hence even with
their addition in the FOF sample, the the results remain qualitatively
similar.

\section{Results}
\subsection{ Angular Momentum of Merger Remnants }

\begin{figure}
\plottwo{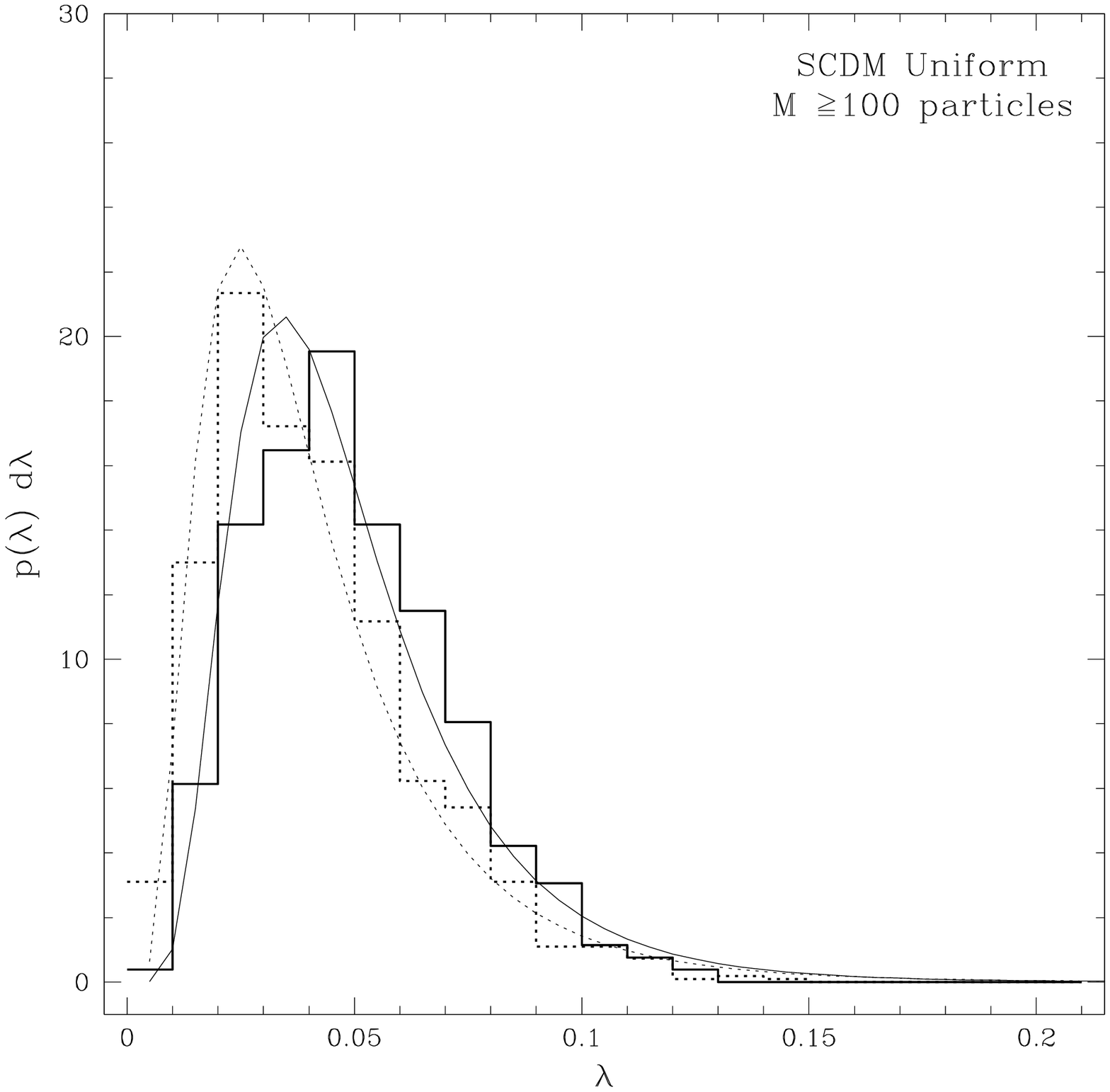}{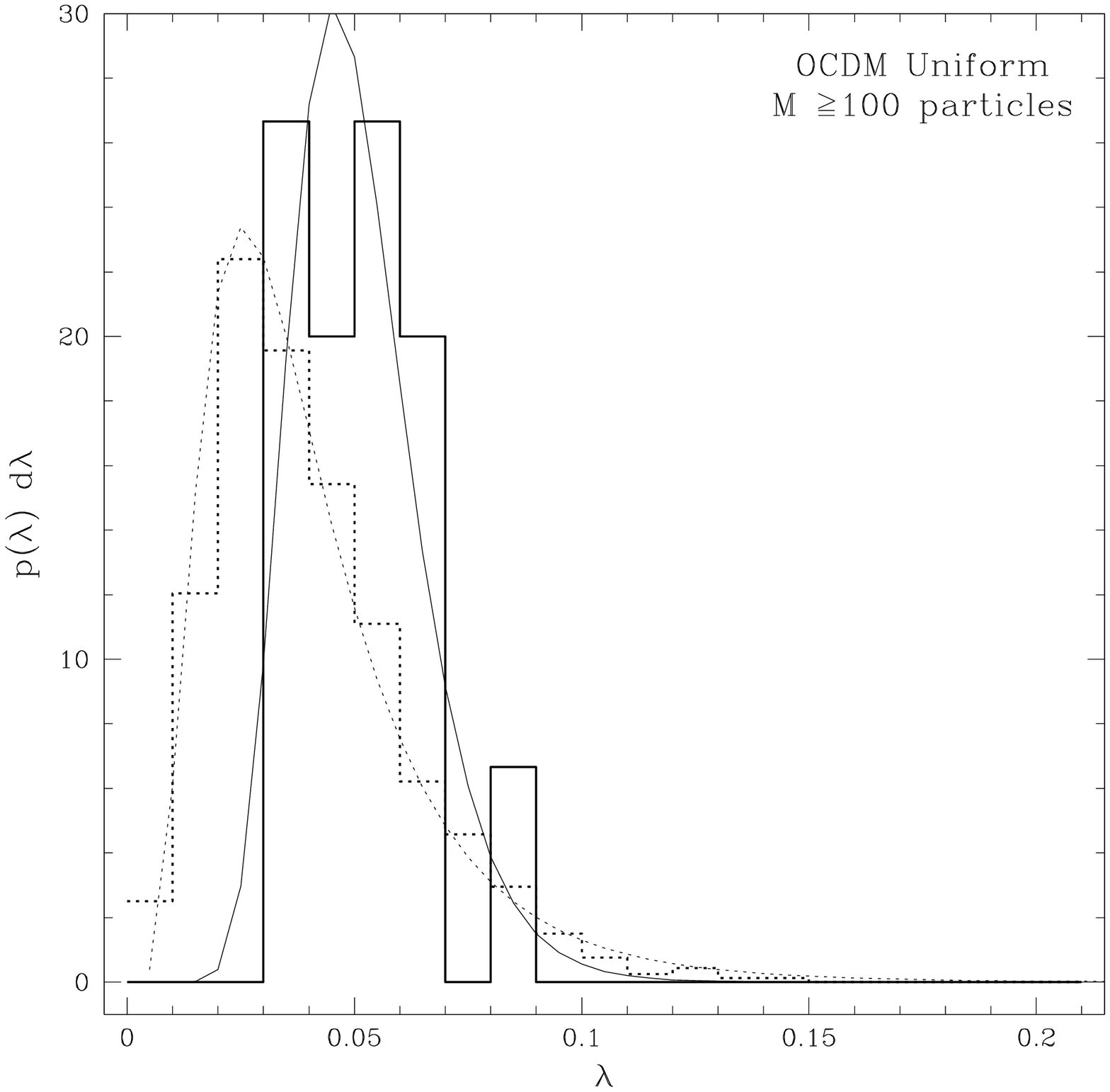}
\plottwo{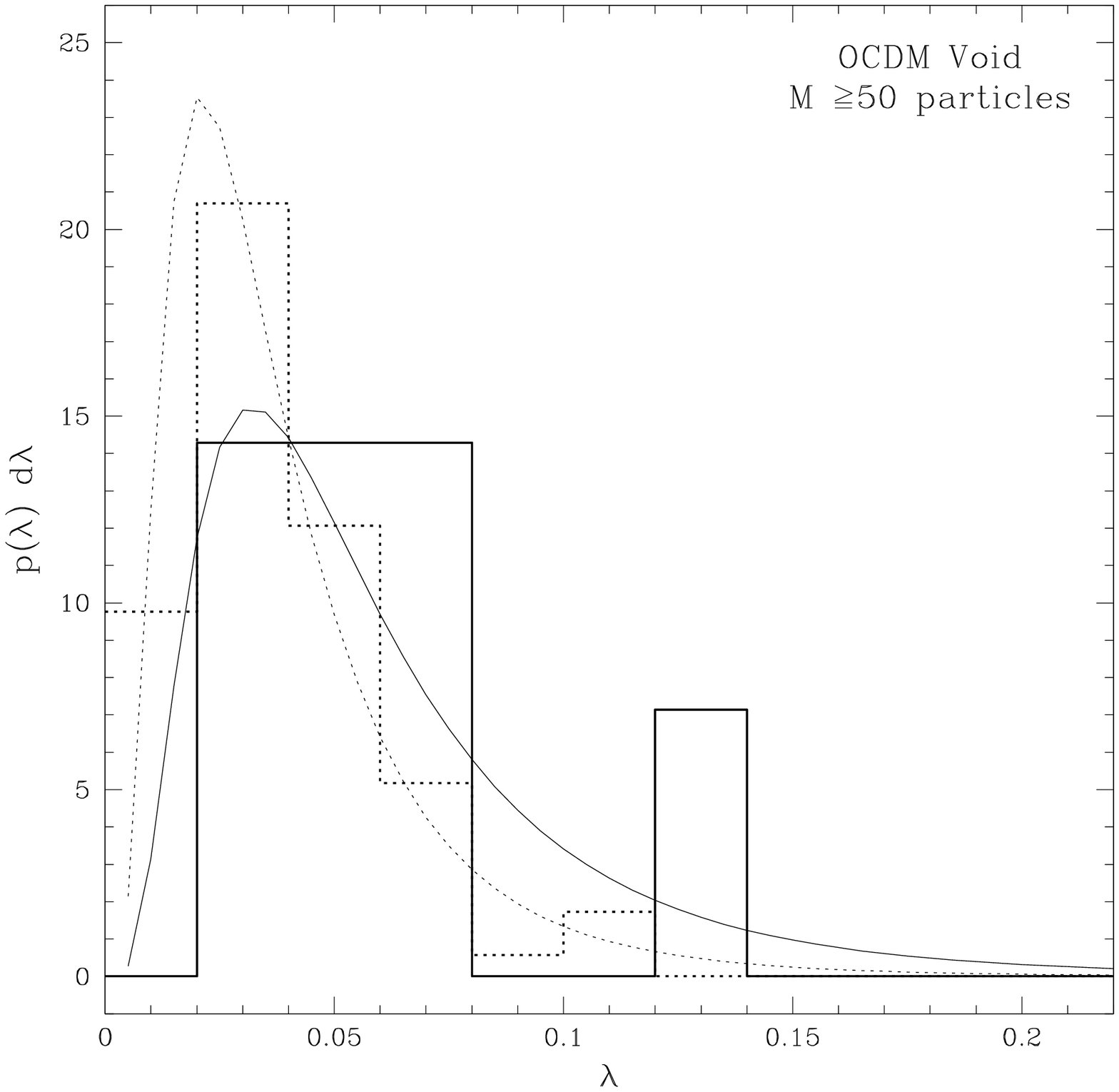}{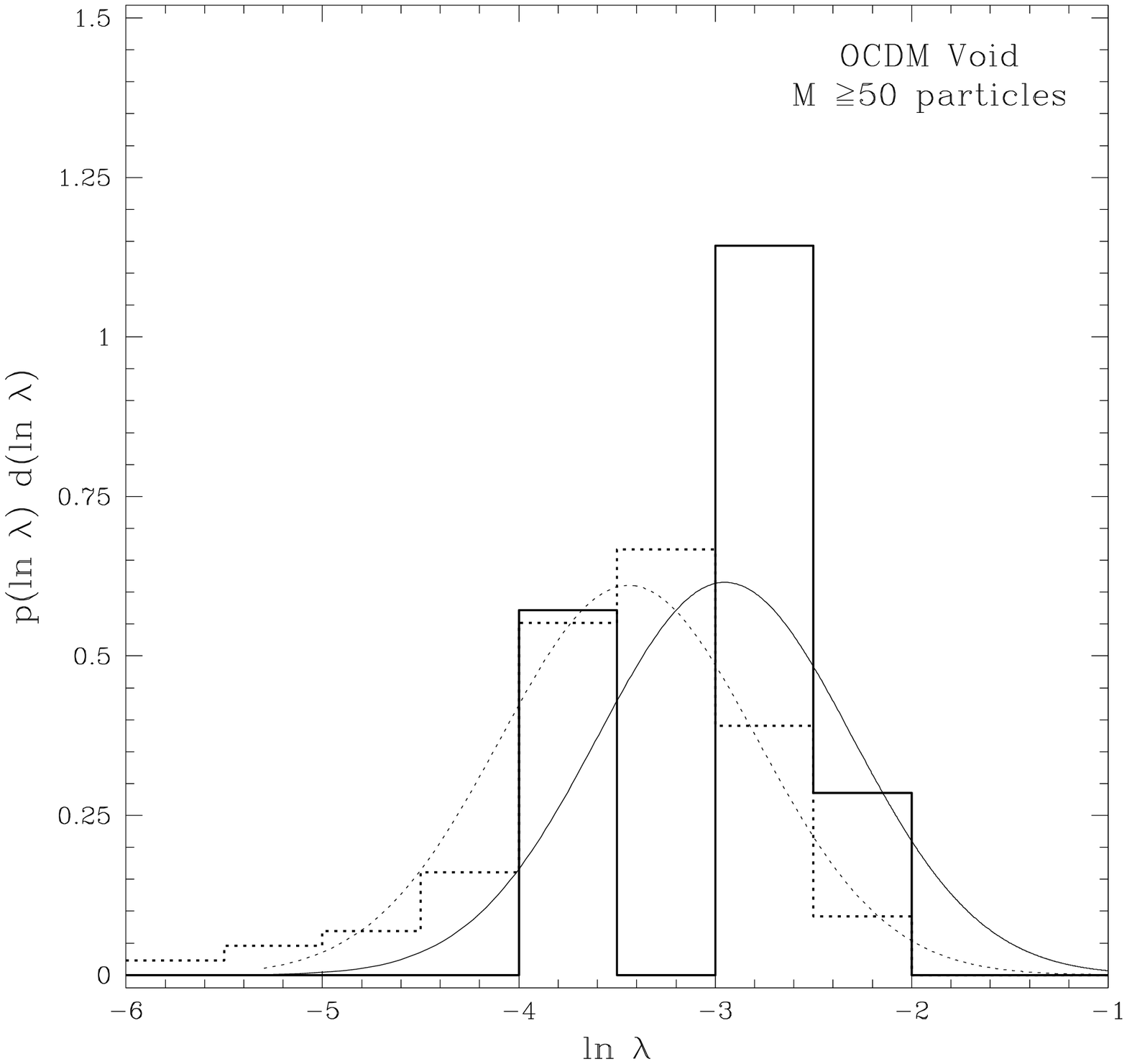}
\caption{Distribution of spin parameters in flat and open uniform
volumes, SCDMu and OCDMu, as well as the open-model void, OCDMv.
Histograms depict the simulated data and curves show the best fit
lognormal function as defined in equation
\protect\ref{eq:lambda}. Dashed lines represent halos which have
experienced no major merger since $z=0.5$ ($z=0.4$ in SCDMu),
whereas solid lines depict the merger remnants.  Since the void run
contains relatively few halos, the spin parameters distribution is
also shown in log space, where it should be Gaussian.
Table~\protect\ref{tab:lambdafits} details fit and distribution
parameters.}
\label{fig:lambda}
\end{figure}

The halo spin parameter distribution in simulations is found to be well approximated by the
lognormal function

\be
\label{eq:lambda}
p(\lambda){\rm d} \lambda = {1 \over \sigma_{\lambda} \sqrt{2 \pi}}
\exp\biggl(- {{\rm ln}^2(\lambda/\bar{\lambda}) \over 2
  \sigma^2_{\lambda}}\biggr) {{\rm d} \lambda \over
  \lambda}
\ee

(\cite{vdbosch98}; \cite{be87}; \cite{ryden88}; \cite{cl96};
\cite{warren92}).  Figure~\ref{fig:lambda} shows $p(\lambda){\rm d}
\lambda$ of simulated halos above $M_{res}$ at z=0.  The solid
histograms denote the merger-remnant population and the dotted
histograms are halos which have not experienced a major merger since
$z=z_m\sim0.5$.  The smooth curves are fits of the form in equation
\ref{eq:lambda} to the simulated distribution.  Parameters for the
fits are given in Table~\ref{tab:lambdafits}.  Since the OCDM void
model contains so few halos, it is easier to see the details of the
distributions by viewing them in log space in
Figure~\ref{fig:lambda}d, where the fitting function is a Gaussian.
It should be noted that Lemson \& Kauffmann (1999) find no
environmental dependence of spin parameter on environment.  Hence, any
differences in OCDMu vs.\ OCDMv arise from the fact that the
simulations sample substantially different ranges in mass and not from
environmental effects.

N-body results are often used to provide an angular momentum
distributions for galaxy halos in semi-analytic models (Dalcanton \ea
1997; Mo \ea 1998; \cite{vdbosch98}).  Warren \ea (1992) find that the
general distribution of halos peaks around $\bar{\lambda} = 0.05$ with
$\sigma_{\lambda} = 0.7$, although later N-body researchers find
distributions more consistent with $0.04 \simlt \bar{\lambda} \simlt
0.05$ and $\sigma_{\lambda} \approx 0.5$ (\cite{cl96}; \cite{mmw98};
see also \cite{sb95}; \cite{ct96}).
I find $\sigma_\lambda$ of all halos (in the uniform volumes) lower
than 0.6 with the merger remnant population tending lower still, while
the peak of the all three simulations presented is $\bar\lambda \sim
0.035$.  Hence, my results for the global halo population are in
concordance with previous findings.

I find that in all cases the distribution of spin parameters of merger
remnants is greater than non-merger remnants.  The K-S probabilities
that the two sets were drawn from the same distribution are given as
$P_{KS}$ in Table~\ref{tab:lambdafits} and are quite low.  The K-S
test is most certain in SCDMv, given the large number of major
mergers.  Even though both OCDM runs contain fewer resolved merger
remnants, the K-S test differentiates the merger and non-merger remnant
populations at substantial confidence.  To determine the robustness of
the dissimilar means $\bar{\lambda}$ of distributions, I use the
Wilcoxon test (\cf Lupton 1993).  The value $P_{Wil}$ given in
Table~\ref{tab:lambdafits} is the probability that the difference
between the mean of the merger population the mean of the non-merger
population could have occurred by chance.  In all cases, the
probability is quite high that $\bar{\lambda}$ of the two
distributions are indeed significantly different.  Consequently, spin
parameters of halos which have experienced a recent major merger are
distributed differently than halos which have not undergone a recent
significant interaction.

Angular momentum has been found to be relatively insensitive to
cosmology and collapse anisotropy for halos formed by less violent
collapse, with the spread in $\bar{\lambda}$ arising mainly from tidal
effects (\cite{huss99}).  This view is supported by Nagashima \& Gouda
(1998) who, using semi-analytics (the ``merging-cell model''), find
that the orbital angular momentum of major mergers is not important
for the remnant, and that major merging activity as a whole has little
effect on the halo angular momentum distribution.  Given the large
spread in $\bar{\lambda}$ in my results, it seems tempting to conclude
that on a halo-by-halo basis, tidal effects are more important than
merger history in determining the ultimate angular momentum of a halo.
Merging may be looked at simply as adding a small upward bias.
However, one should note that the dispersion $\sigma_\lambda$ in the
merger remnants is typically lower than the non-merger remnants.  This
can imply one of two possibilities: A) the angular momentum of mergers
is largely independent of tidal torques, depending more on the mergers
themselves or B) merging adds selects for specific environments.  In
B), by considering mergers, we are simply biasing ourselves toward
examining halos which are in a region that promotes merging, nurturing
halos with a narrower range in $\lambda$ and a higher mean.  In A), it
is a combination of the orbital angular momentum of the merger and the
spin-alignment of the halos that determines a remnant's resultant
angular momentum.  However, if the spin-alignment were dominant, one
would expect the dispersion in $\lambda$ to be {\em larger} than the
non-merger distribution, as two average halos can either add their
spins or have them cancel completely.  Consequently in case A), the
orbital angular momentum of merging halos is the more likely
contributor to the final remnant momentum.  To summarize, tidal
effects can dominate the merger remnant angular momentum distribution
only if the lower dispersion of the merger remnant population can be
explained.  A likelier explanation is that by examining mergers, one
is merely selecting a specific environment and hence only a subset of
the possible tidal torques.  The equally likely alternative is that
the orbital angular momentum of the pre-merger halos dominates the
spin parameters of merger remnants.
 
\begin{figure}
\plotone{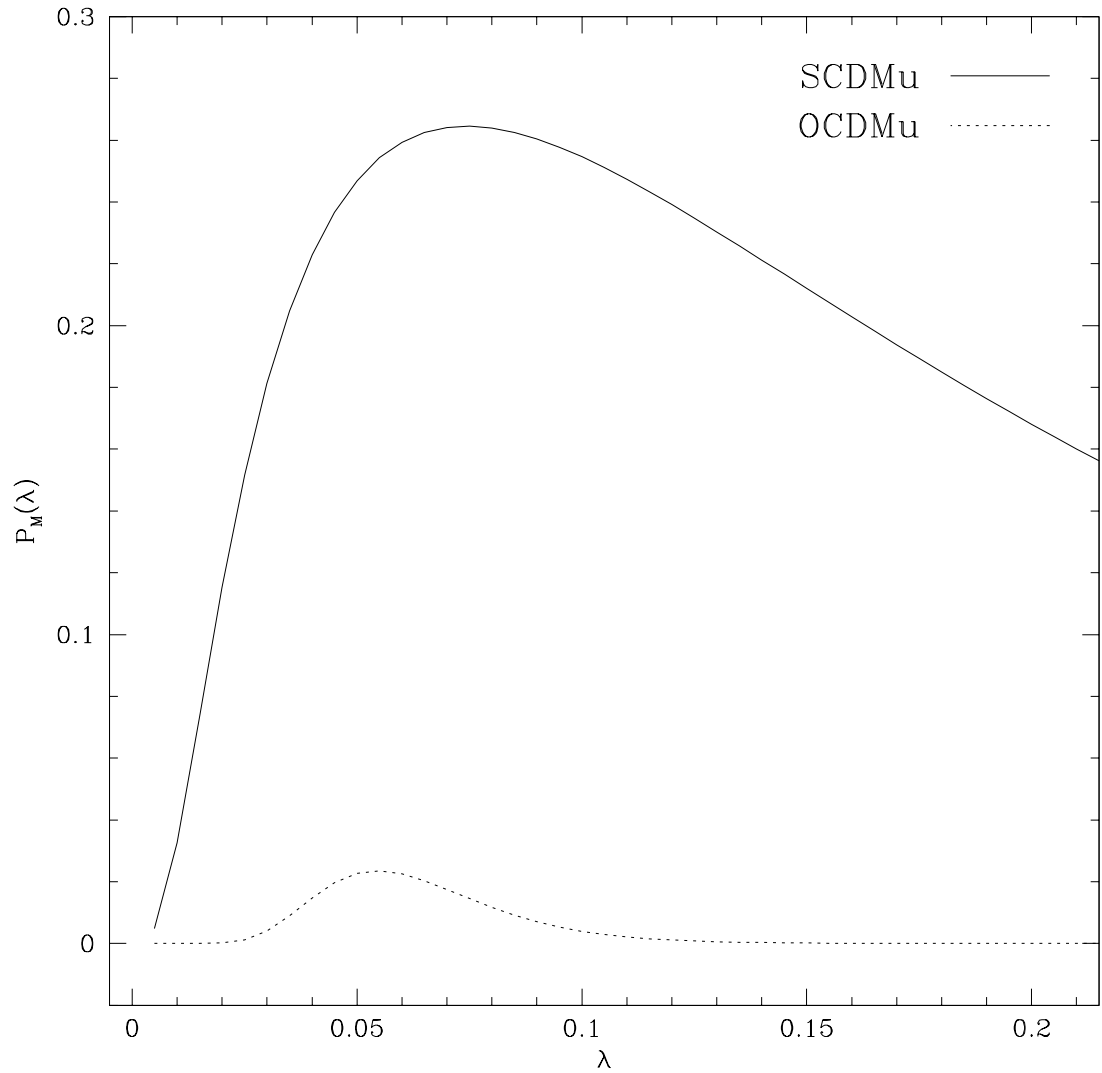}
\caption{The probability $P_M$ that a halo with a given spin parameter
$\lambda$ is a merger remnant for SCDMu (solid line) and OCDMu (dotted
line).}
\label{fig:lamdiff}
\end{figure}

How relevant is the contribution of mergers to the overall spin
parameter distribution?  Figure~\ref{fig:lamdiff} plots the fraction
of halos in my sample with spin parameters $\lambda$ that are major
merger remnants.  The result is strongly dependent on cosmology, given
that in open universes, merging activity at low redshift is
substantially less likely than in $\Omega_0=1$ models
(\cite{mergerpaper}).  Consequently, the global effect of merging on
halo angular momentum is simply determined by the rate of major
mergers.  This information is, however, of marginal utility since
galaxy history or properties are rarely used to infer angular
momentum.  In general, the distribution $p(\lambda) d\lambda$ is
employed in an {\it a priori} manner.  In this case, researchers may
wish to adopt a different distribution for galaxies which are merger
remnants and those which are not.

The overall implication of these results is that merging activity is
correlated with higher angular momentum.  Orbital angular momentum of
interaction halos would seem to be the cause most consistent my data.
However, on galaxy scales, the angular momentum of ellipticals is
generally much lower than spirals (\cite{ng98}).  This is troublesome
if ellipticals are to be considered merger remnants and spirals the
results of long periods of quiescence.  The quiescent accretion
required to form a higher-$\lambda$ spiral disk would appear to take
place in halos with systematically lower spin parameters.  Differences
in angular momenta between galaxy morphologies may be more due to
formation epoch (\cite{huss99}) than merger history.

\begin{table}
\begin{center}
\begin{tabular}{crllcrllcrllll}
\multicolumn{1}{c}{ }&\multicolumn{3}{c}{All Halos}&\multicolumn{1}{c}{ }&
\multicolumn{3}{c}{Merger Remnants}& \multicolumn{1}{c}{ }&
\multicolumn{3}{c}{Non-Merger Remnants} \\ \cline{2-4} \cline{6-8} \cline{10-12}
\multicolumn{1}{c}{Model}&\multicolumn{1}{c}{$N$}&\multicolumn{1}{c}{$\bar\lambda$}&
\multicolumn{1}{c}{$\sigma_\lambda$}&\multicolumn{1}{c}{ }&
\multicolumn{1}{c}{$N$}&\multicolumn{1}{c}{$\bar\lambda$}&\multicolumn{1}{c}{$\sigma_\lambda$}&
\multicolumn{1}{c}{ }&\multicolumn{1}{c}{$N$}&\multicolumn{1}{c}{$\bar\lambda$}&
\multicolumn{1}{c}{$\sigma_\lambda$}&\multicolumn{1}{c}{$P_{KS}$}&
\multicolumn{1}{c}{$P_{Wil}$} \\
\tableline
SCDMu& 1353& 0.0366& 0.585&& 261& 0.0438& 0.500&& 1092& 0.0351&
0.596& $10^{-7}$& $10^{-7}$\\
OCDMu& 1609& 0.0353& 0.568&&  15& 0.0492& 0.278&& 1594& 0.0352&
0.569& 0.0119& $10^{-3}$\\
OCDMv&      94& 0.0331& 0.660&&   7& 0.0495& 0.657&&   87& 0.0320&
0.653& 0.320& 0.132\\
\tableline
\end{tabular}
\end{center}
\caption{Details of the halo spin parameter distribution for the
sample of all halos, merger remnant halos, and non-merger remnant
halos.  $N$ denotes the number of halos in
each distribution.  The fitting parameters $\bar\lambda$ and
$\sigma_\lambda$ for the lognormal distribution $p(\lambda) d\lambda$
are defined in equation \protect\ref{eq:lambda}.   $P_{KS}$ is the K-S
probability that the merger and non-merger populations were drawn from
the same distribution.  $P_{Wil}$ is the Wilcoxon probability that
$\bar\lambda$ of the merger and non-merger populations occurred by chance.}
\label{tab:lambdafits}
\end{table}

\subsection{Halo Mass Histories and Formation Times}

\begin{figure}
\plotone{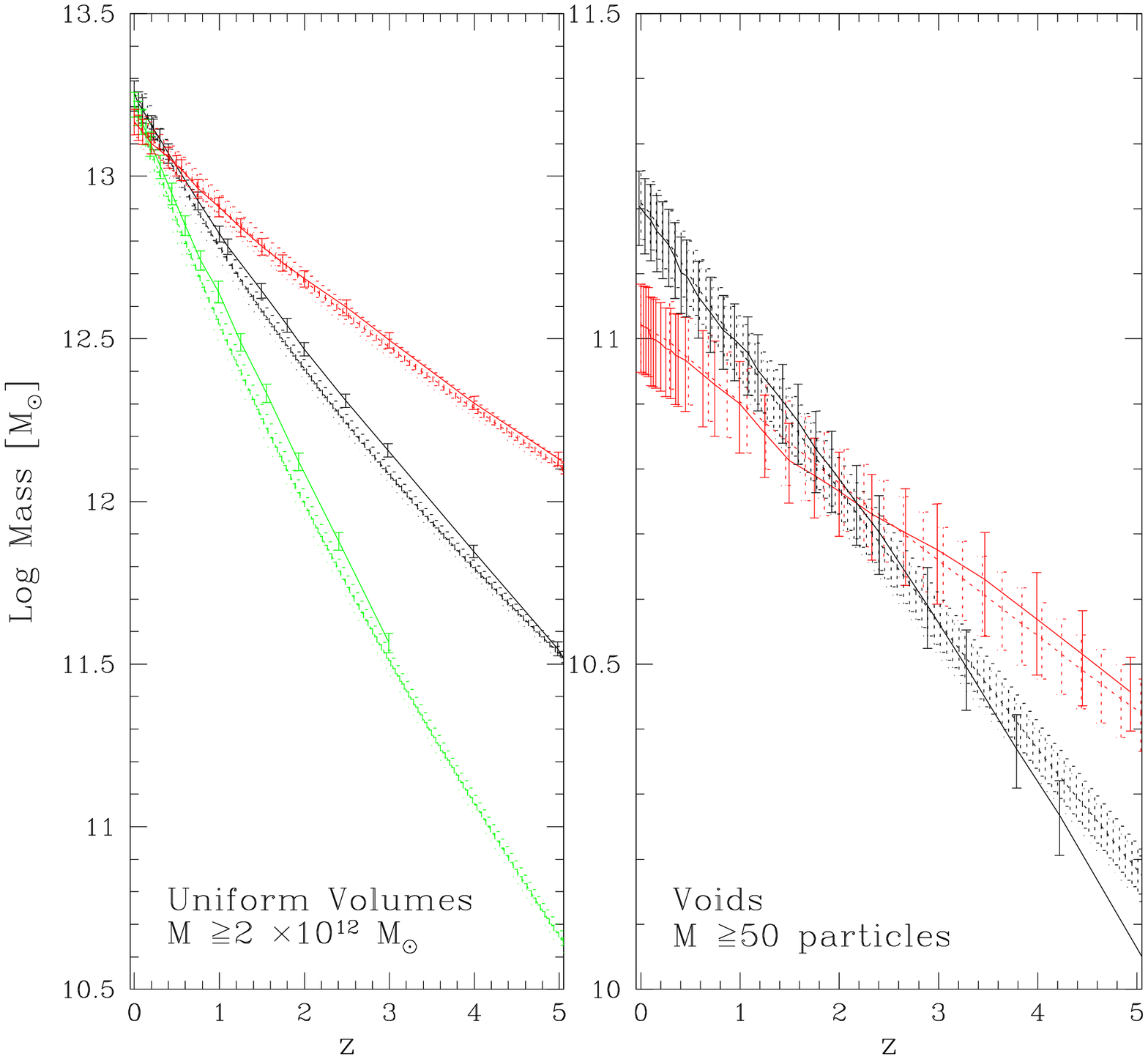}
\caption{Average most massive progenitor mass vs.\ redshift $z$.  The
solid curves denote the average mass of the most massive progenitor of
all halos in the simulations greater than the mass cut $M \geq 2\times
10^{12} M_\odot$ (uniform volumes, left panel) or $M >$ 50 particles
(voids, right panel).  The dashed curves are the Monte-Carlo realizations of
the same sample of halos.  The errorbars for all curves at the
$1-\sigma$ deviations of 1000 bootstrap samples of the $z=0$ halo
sample.  In the left panel, each pair of curves (\ie one dashed and
one solid) represents a different cosmology from top to bottom view at
$z=5$: OCDMu, SCDMu, TCDMu.  In the right panel, the steep pair of
curves denotes the SCDMv run, and the shallower the OCDMv run.}
\label{fig:masshist}
\end{figure}

Here I compare the Monte-Carlo approximation to simulated data
in differing cosmologies and environment.  Figure~\ref{fig:masshist}
shows the mean mass histories of halos selected in accordance with
section \ref{ssec:masshistmethod} for the uniform volumes in all three
cosmologies, and the voids in SCDM and OCDM.  This is essentially a
measurement of the conditional mass function examined by other authors
(\cite{tormen98}; \cite{somerville00}; \cite{bower91}).
It has been established in previous studies that the Press-Schechter
model underestimates the number of halos relative to simulations at
mass scales which lessen with increasing redshift (\cite{somerville00}, \cite{gross98}, \cite{tormen98}).  Somerville \ea (2000) find that for their tilted CDM
model, EPS underpredicts the masses of the most massive progenitor at
redshifts $z \simgt 0.5$.  In the flat uniform volumes, I also find that
EPS systematically predicts lower-mass principle
progenitors than simulations.  The discrepancy in the average
progenitor mass is as high as 20\% in SCDM and TCDM.  
Interestingly enough, things are different in the OCDMu open model in
contrast to the findings of Somerville \ea (2000).  In this case, the
semi-analytic predictions appear to be roughly in line with the simulations.
Also, EPS appears to reproduce the mean mass histories of
both voids reasonably well, trailing outside the 1-$\sigma$ dispersion
bars for SCDMv only at very high redshift.  It should be noted that
the void simulations sample halos of a much lower mass range resulting
in higher typical formation redshifts and making a direct comparison
between uniform volume and void runs somewhat hazardous.  The
difference in mass history between SCDMu vs.\ SCDMv and OCDMu vs.\
OCDMv is most likely caused by the difference in mass range.  The important comparison is between Monte-Carlo and
simulated results in each regime.  The masses of the halos in the void
runs are somewhat lower than in the uniform volumes and thus farther
from $M_*$ of the mass function.  Since the shape of the mass function
changes little in this low-mass regime, it may be possible for
$10^{11} M_\odot$ halos to match while $10^{13} M_\odot$ halos do not.
Somerville \ea (2000) also find that EPS progenitor masses match better
for halos of mass $\sim 10^{12} M\odot$ than for halos $M > 10^{13}
M_\odot$.  Hence, these results are reasonably consistent with their
findings.  Given the small number of halos in the void, however, these
results are not conclusive and should be reexamined with a larger
sample size.

\begin{figure}
\plotone{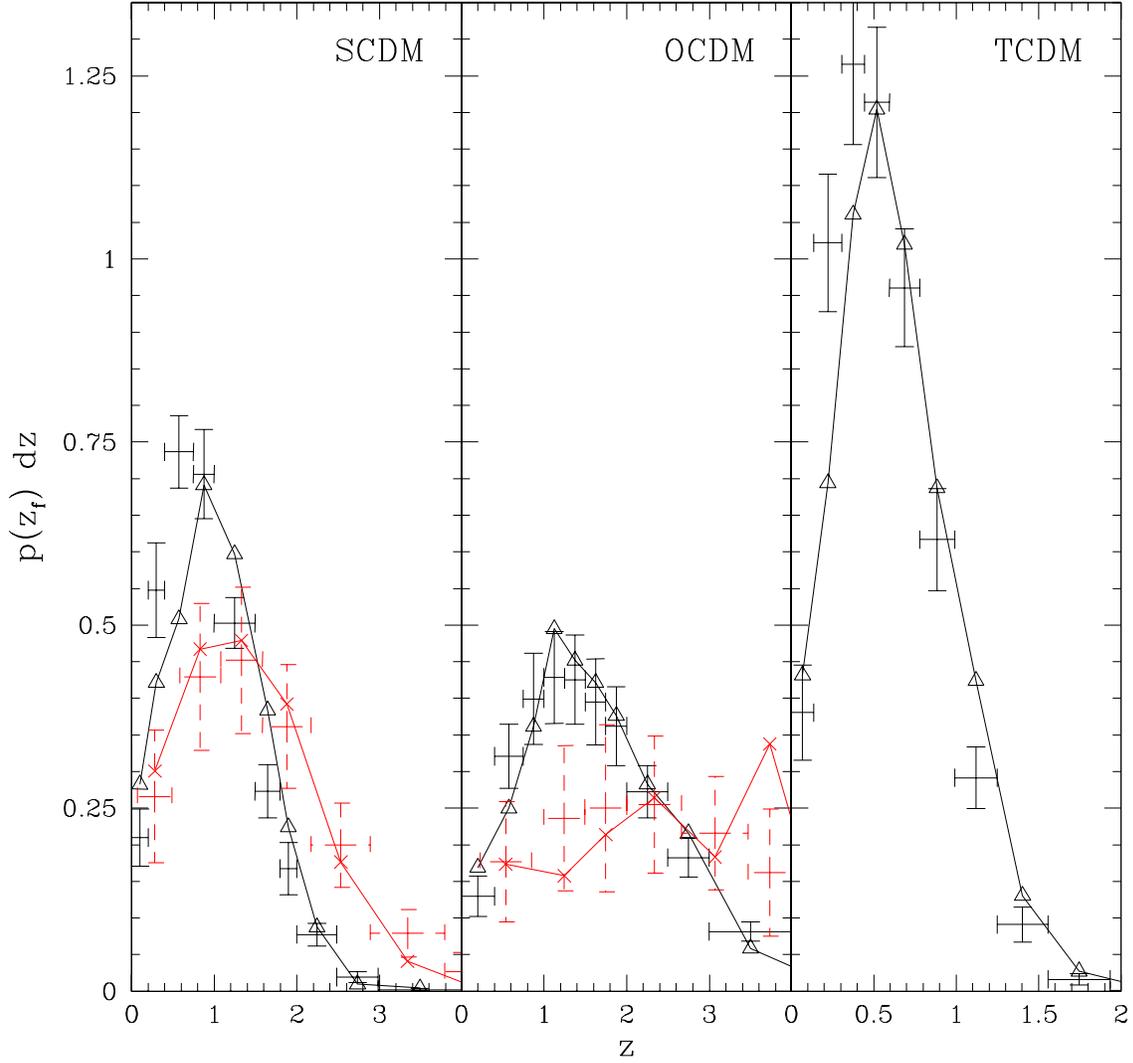}
\caption{Probability distribution of formation redshift $p(z_f) dz$
vs.\ redshift $z$.  In each model, the solid line and triangular
points represent the uniform volume, while the solid line and starred
points denote the void simulations.  The Monte-Carlo results are
indicated by errorcrosses, with solid representing the uniform volumes
and dashed representing the voids.  The horizontal errorcross
components denote the bin boundaries which were used for both the
Monte-Carlo and simulated halos.  The vertical components enclose 90\%
of 1000 bootstrap resamplings of the Monte-Carlo realizations.  Thus, for EPS
to match the simulated formation times the vertical errorbars must
enclose 9 out of 10 points.  Note that the horizontal scale for TCDM
is half that of the other panels.}
\label{fig:formtime}
\end{figure}

The difference in results can be examined more quantitatively by
comparing halo formation times.  A halo is said to have formed when
the mass of the most massive progenitor is 50\% of the z=0 mass.
Given the discreteness of the simulation outputs, the formation time
of simulated halos can only be bounded between adjacent outputs where
the halo transitions the 50\% formation threshold.
Figure~\ref{fig:formtime} plots the formation times of halos in
simulations vs. their respective Monte-Carlo realizations.  The points
from the simulations (connected by lines) denote the number of halos
formed in a bin $\Delta z$ where the bin boundaries correspond to
simulation output times.  In some cases multiple simulation outputs
are combined into a single bin in order to increase statistics.  In
order to make a meaningful comparison, the formation time for each
Monte-Carlo realization is binned in exactly the same manner as the
simulation output.  The bin boundaries are given by the horizontal
bars in the Monte-Carlo data in Figure~\ref{fig:formtime}.  To
determine confidence limits, I bin 1000 bootstrap samples of the
Monte-Carlo realization and show the regime in which 90\% of the
realizations fall as vertical errorbars.  Consequently, for the
simulation to match the EPS prediction, 90\% of the simulation points
should fall within the Monte-Carlo errorbars.  As we expect from the
mass history results, the EPS results are clearly different from SCDMu
and TCDMu simulations at the 90\% confidence level.  The median
formation time is a good measure for the ``error'' in formation time
between simulated and EPS realizations.  In SCDMu, the median
simulated formation time in units of the $z=0$ age of an SCDM universe
is approximately 0.20 while the Monte-Carlo realizations find a median
formation time of 0.38.  For OCDMu, 4 out of 10 simulated points are
outside the errorbars, meaning that EPS also fails to match at the
90\% confidence level.  In the case of the void runs, the sampling
errors are too large to rule out either Monte-Carlo realization,
consistent with the interpretation from Figure~\ref{fig:masshist}.
Consequently, the EPS appears to diverge notably from simulated
realizations in flat models, as well as maintain a measurable
difference in open models.  The difference in the results between open
and flat-matter cosmologies is most likely related to the growth of
fluctuations in each.  Since structure grows faster in $\Omega_0=1$
models, the differences between EPS and simulations is most likely
exaggerated with respect to low-$\Omega_0$.

Note again that these results allow no freedom of choice
of the critical density parameter $\delta_c$ as the PS and simulated
mass functions are constrained to match at redshift $z=0$.  
Governato \ea (1999) provide evidence that a better fit can be
attained by evolving $\delta_c$ in a nonlinear fashion.  Hence,
one is effectively renormalizing the mass function at every epoch,
thus changing the mapping between perturbation spectrum and mass.  If I use
their fitting formula $\delta_c(z)=1.686\left[ (0.7/\sigma_8)(1+z)
\right]^{-0.125}$ (valid for $\Omega_0=1$) for SCDMu, I find that this
does indeed change the Monte-Carlo mass history of Figure~\ref{fig:masshist}b
substantially.  While the Monte-Carlo history is lower for small redshifts, it
crosses the simulated curve and eventually diverges $2\sigma$ above the
simulated histories for $z>1.4$.  Governato \ea construct this
relation from cluster-mass halos, well over an order of magnitude more
massive than this study, so it is not surprising that using this fit
does not induce EPS to match my simulated data.  Still, the
simulated results are bracketed with static $\delta_c$ producing too rapid
halo growth and redshift-variant $\delta_c$ producing halos which grow
slower than in N-body simulations.  It may be possible to fine-tune a
prescription for $\delta_c(z)$ to simulated data.  As with any fine
tuning of semi-analytic parameters, however, it is essential to
understand the physical mechanisms underlying such parameters lest
their predictive power be compromised.  A study of such scope is best
left to a future paper.

\section{Conclusions}

I present results from numerical N-body simulations regarding the
effect of merging events on the angular momentum distribution of
galactic halos as well as a comparison of halo growth in Monte-Carlo
vs.\ N-body methods.  In all three models tested and both in a uniform
sample and in an underdense region, the distribution of halo spin
parameters of merger remnants is greater than non-merger remnants.
The spin parameters of the global halo population are consistent with
other recent results \cite{cl96}; \cite{mmw98}; \cite{sb95};
\cite{ct96}).  However, the K-S probabilities of merger and non-merger
samples being from the same sample are quite low, and the mean spin
parameters of the two samples $\bar{\lambda}$ are significantly
different.  The actual contribution of merging on the global spin
parameter distribution depends on the fraction of halos that are
merger-remnants and hence is greatly affected by cosmology.  The
overall effect of merging is to increase the mean and decrease the
dispersion of the spin parameter distribution.  I offer the tentative
conclusion that this is consistent with the orbital angular momentum
of the merging halos playing a major role in establishing the
remnant's resultant net angular momentum.  The causes and
time-evolution of this behavior should be investigated in greater
detail in future studies using simulations of superior dynamic range.
It is also possible that angular momentum distributions could vary as
a function of redshift, but I leave this for a future paper.

I trace the most massive progenitor of halos of mass $M_i \sim 10^{13}
M_\odot$ at redshift $z=0$ back to $z=5$ for ``uniform volumes'' in
all three cosmologies as well as mass $M_i \sim 10^{11} M_\odot$ halos
for large voids in the $\Omega_0=1$ and $\Omega_0=0.3$ cosmologies.  I
find that in the uniform volumes, the Press-Schechter based method
reproduces the general trend of the simulated mass histories
relatively well, although in $\Omega_0=1$ models it underpredicts the
average mass history of halos by roughly 20\%.  This leads to a
further discrepancy in the estimation of halo formation times.  For
example, the median formation time of halos in a flat universe (SCDMu)
is 20\% the age of the Universe according to simulations, but 38\% the
age of the Universe in Monte-Carlo realizations. 
Because the PS $z=0$ mass function is constrained to match the simulated
halo mass function, $\delta_c$ is no longer a free parameter.  It may
be possible fine-tune PS to match simulated results by the use of a
$\delta_c$ which evolves non-linearly with redshift, although such a
study is beyond the scope of this paper.

My results are consistent with work by Somerville \ea (2000) and
Tormen (1998).  Somerville \ea find that EPS reproduces relative
properties of the progenitor halo distribution, \eg their mass ratios,
quite well.  However, the {\em absolute} progenitor masses and the
overall conditional mass function exhibits discrepancies of up to
50\%.  Tormen concludes that the EPS formulation is inadequate to
describe clustering in a constrained environment and may possibly be
inadequate in a general context. I find that the latter is indeed
true, as the conditional halo mass function in a cosmologically
representative sample does not agree with simulated data.
Consequently, care should be taken when comparing the raw masses of
halo progenitors, or their mass relative to their eventual
``children'' of the present day.  Formation times of halos from the
EPS conditional mass function are systematically biased.  Furthermore,
care should be taken when examining the redshift evolution of any
absolute quantity which depends on the evolution of mass, such
luminosity or color.  Halo properties should be compared in a relative
sense at each redshift and not in an absolute sense with respect to
their parents or children at other times.

Recently, alternative expressions for the PS mass function have been
proposed and investigated.  Sheth \& Tormen (1999) modify the
Press-Schechter formalism, producing a much better match to the mass
function, especially at the low-mass end, of simulations.  Mo, Sheth,
\& Tormen (1999) have shown that significant improvements to the
standard PS approach can be made by modeling the formation of bound
structures as an ellipsoidal rather than a spherical collapse.
Jenkins \ea (2000) derive an empirical mass function by fitting
the results numerous N-body simulations.  My results further
demonstrate the need for using these alternative approaches when
generating Monte-Carlo mass histories.

\acknowledgements

The author gratefully acknowledges the generosity of Paolo Tozzi for
providing his semi-analytic code for this study.  The author also
thanks Fabio Governato, without whose guidance and support this paper
would have been impossible, and Tom Quinn who never ceases to provide
useful advice and feedback.  Furthermore, it is a pleasure to
acknowledge frequent and useful discussions with Frank van den Bosch
and Julianne Dalcanton.  The author was supported by NASA Grant
NGT5-50078 for the duration of this work.  The simulations were
performed on the Cray T3D/E at the Arctic Region Supercomputing Center.

\end{document}